\documentclass[twocolumn]{aastex61}

\newcommand\aastex{AAS\TeX}

\shorttitle{\aastex\ Axisymmetric Boltzmann simulations of CCSNe}
\shortauthors{Nagakura et al.}

\begin{document}
\title{Simulations of Core-Collapse Supernovae in spatial axisymmetry with Full Boltzmann Neutrino Transport}
\correspondingauthor{Hiroki Nagakura}
\email{hirokin@caltech.edu}

%% \author{Hiroki Nagakura$^{1}$, Wakana Iwakami$^{2,3}$, Shun Furusawa$^{4}$, Hirotada Okawa$^{2,3}$, Akira Harada$^{5}$, \\
%% Kohsuke Sumiyoshi$^{6}$, Shoichi Yamada$^{3,7}$, Hideo Matsufuru$^{8}$ and Akira Imakura$^{9}$
%% }

%% \affiliation{$^1$TAPIR, Walter Burke Institute for Theoretical Physics, Mailcode 350-17, California Institute of Technology, Pasadena, CA 91125, USA}

%% \affiliation{$^2$Yukawa Institute for Theoretical Physics, Kyoto
%%   University, Oiwake-cho, Kitashirakawa, Sakyo-ku, Kyoto, 606-8502,
%%   Japan}

%% \affiliation{$^3$Advanced Research Institute for Science \&
%% Engineering, Waseda University, 3-4-1 Okubo,
%% Shinjuku, Tokyo 169-8555, Japan}

%% \affiliation{$^4$ Frankfurt Institute for Advanced Studies, J.W.Goethe University, D-60438 Frankfurt am Main, Germany}

%% \affiliation{$^5$ Department of Physics, University of Tokyo, 7-3-1 Hongo, Bunkyo, Tokyo 113-0033, Japan}

%% \affiliation{$^6$Numazu College of Technology, Ooka 3600, Numazu, Shizuoka 410-8501, Japan}

%% \affiliation{$^7$Department of Science and Engineering, Waseda
%%   University, 3-4-1 Okubo, Shinjuku, Tokyo 169-8555, Japan}

%% \affiliation{$^8$High Energy Accelerator Research Organization, 1-1 Oho, Tsukuba, Ibaraki 308-0801, Japan}

%% \affiliation{$^9$University of Tsukuba, 1-1-1, Tennodai Tsukuba, Ibaraki 305-8577, Japan}

\author{Hiroki Nagakura}
\affiliation{TAPIR, Walter Burke Institute for Theoretical Physics, Mailcode 350-17, California Institute of Technology, Pasadena, CA 91125, USA}

\author{Wakana Iwakami}
\affiliation{Yukawa Institute for Theoretical Physics, Kyoto
University, Oiwake-cho, Kitashirakawa, Sakyo-ku, Kyoto, 606-8502,
Japan}
\affiliation{Advanced Research Institute for Science \&
Engineering, Waseda University, 3-4-1 Okubo,
Shinjuku, Tokyo 169-8555, Japan}

\author{Shun Furusawa}
\affiliation{Interdisciplinary Theoretical Science (iTHES) Research Group,
RIKEN, Wako, Saitama 351-0198, Japan}

\author{Hirotada Okawa}
\affiliation{Yukawa Institute for Theoretical Physics, Kyoto
University, Oiwake-cho, Kitashirakawa, Sakyo-ku, Kyoto, 606-8502,
Japan}
\affiliation{Advanced Research Institute for Science \&
Engineering, Waseda University, 3-4-1 Okubo,
Shinjuku, Tokyo 169-8555, Japan}

\author{Akira Harada}
\affiliation{Department of Physics, University of Tokyo, 7-3-1 Hongo, Bunkyo, Tokyo 113-0033, Japan}

\author{Kohsuke Sumiyoshi}
\affiliation{Numazu College of Technology, Ooka 3600, Numazu, Shizuoka 410-8501, Japan}

\author{Shoichi Yamada}
\affiliation{Advanced Research Institute for Science \&
Engineering, Waseda University, 3-4-1 Okubo,
Shinjuku, Tokyo 169-8555, Japan}
\affiliation{Department of Science and Engineering, Waseda University, 3-4-1 Okubo, Shinjuku, Tokyo 169-8555, Japan}

\author{Hideo Matsufuru}
\affiliation{High Energy Accelerator Research Organization, 1-1 Oho, Tsukuba, Ibaraki 305-0801, Japan}

\author{Akira Imakura}
\affiliation{University of Tsukuba, 1-1-1, Tennodai Tsukuba, Ibaraki 305-8577, Japan}

%In the early post-bounce phase, the prompt convection is developed for both models while the LS EOS is slightly more vigorous than the FS EOS. We find that 

% Despite of the difference, the time evolution of average shock radius between two models are very similar up to $\sim 60$ms after bounce, and then deviate from 

\begin{abstract}
%We present results of spatially axisymmetric core-collapse supernova simulations with full Boltzmann neutrino transport, which amount to a time-dependent 5-dimensional (2 in space and 3 in momentum space) problem in fact.
We present the first results of our spatially axisymmetric core-collapse supernova simulations with full Boltzmann neutrino transport,   which amount to a time-dependent 5-dimensional (2 in space and 3 in momentum space) problem in fact. Special relativistic effects are fully taken into account with a two-energy-grid technique. We performed two simulations for a progenitor of $11.2 M_{\sun}$, employing different nuclear equations-of-state (EOS's): Lattimer and Swesty's EOS with the incompressibility of $K=220$MeV (LS EOS) and Furusawa's EOS based on the relativistic mean field theory with the TM1 parameter set (FS EOS). In the LS EOS the shock wave reaches $\sim 700{\rm km}$ at $300{\rm ms}$ after bounce and is still expanding whereas in the FS EOS it stalled at $\sim 200{\rm km}$ and has started to recede by the same time. This seems to be due to more vigorous turbulent motions in the former during the entire post-bounce phase, which leads to higher neutrino-heating efficiency in the neutrino-driven convection. We also look into the neutrino distributions in momentum space, which is the advantage of the Boltzmann transport over other approximate methods. We find non-axisymmetric angular distributions with respect to the local radial direction, which also generate off-diagonal components of the Eddington tensor. We find that the $r {\theta}$-component reaches $\sim 10 \%$ of the dominant rr-component and, more importantly, it dictates the evolution of lateral neutrino fluxes, dominating over the $\theta \theta$-component, in the semi-transparent region. These data will be useful to further test and possibly improve the prescriptions used in the approximate methods.
\end{abstract}




\section{Introduction} \label{sec:intro}

The theoretical study of the explosion mechanism
of core-collapse supernovae (CCSNe)
 has heavily relied on numerical simulations. This is mainly 
because nearby CCSNe are rare~\citep{1991ARA&A..29..363V,1993A&A...273..383C,1994ApJS...92..487T,2005AJ....130.1652R,2006Natur.439...45D,2010MNRAS.407.1314M,2011MNRAS.412.1441L} and, in fact, SN1987A is 
the only one close enough to extract some useful information on what happened deep inside the massive star 
from, among other things, the detection of neutrinos~\citep{1987PhRvL..58.1494B,1987PhRvL..58.1490H}.
Since the CCSNe are intrinsically multi-scale, multi-physics and multi-dimensional (multi-D) phenomena, their mechanism can be addressed only with detailed numerical simulations.

%2008ApJ...685.1069O,

Unfortunately, even the most advanced multi-D simulations of CCSNe
%, which produced successful explosions rather casually these days,
employed approximations one way or another in their numerical treatment of neutrino transport~\citep{2009ApJ...694..664M,2012ApJ...756...84M,2013ApJ...767L...6B,2014ApJ...786...83T,2016ApJ...818..123B,2015ApJ...800...10D,2015ApJ...807L..31L,2015ApJ...801L..24M,2016ApJS..222...20K,2016ApJ...831...81S,2015arXiv151107443O,2016ApJ...817...72P,2015MNRAS.453.3386J,2016ApJ...825....6S,2016ApJ...831...98R,2017MNRAS.468.2032A,2016arXiv161105859B}.
%~\citep{2012ApJ...756...84M,2013ApJ...767L...6B,2014ApJ...786...83T,2016ApJ...818..123B,2015ApJ...800...10D,2015ApJ...807L..31L,2015ApJ...801L..24M,2016ApJS..222...20K,2016ApJ...831...81S,2015arXiv151107443O,2016ApJ...817...72P,2015MNRAS.453.3386J,2016ApJ...825....6S,2016ApJ...831...98R,2017MNRAS.468.2032A,2016arXiv161105859B}.
% Most of them somehow integrated out the angular degrees of freedom in momentum space~\citep{2015ApJ...800...10D,2016ApJS..222...20K,2016ApJ...831...81S,2015arXiv151107443O,2015MNRAS.453.3386J,2016ApJ...831...98R,2016arXiv161105859B} or neglected non-radial fluxes in neutrino transport~\citep{2012ApJ...756...84M,2013ApJ...767L...6B,2014ApJ...786...83T,2016ApJ...818..123B,2015ApJ...807L..31L,2015ApJ...801L..24M,2016ApJ...831...81S,2016ApJ...817...72P,2016ApJ...825....6S,2017MNRAS.468.2032A}, and some of these computations seem to be at odds with each one another probably as a consequence of these approximations.
 Most of them somehow integrated out the angular degrees of freedom in momentum space or neglected non-radial fluxes in neutrino transport.
% and some of these computations seem to be at odds with one another probably as a consequence of these approximations.
\citet{2008ApJ...685.1069O} is the only exception, in which they conducted time-dependent 5-dimensional simulations in spatial axisymmetry.
However, they ignored relativistic corrections completely, dropping all fluid-velocity-dependent terms, which are crucial for qualitatively correct descriptions of the angular distribution of neutrinos in momentum space (see e.g.,~\citet{2006A&A...447.1049B,2012ApJ...747...73L}).

 The best way to calibrate all these approximate methods should be to compare them with simulations that solve full Boltzmann equations, retaining the angular degree of freedom, for neutrino transport. Under axisymmetry in space, this is possible now indeed and we have achieved such simulations with the K computer in Japan, one of the currently available best supercomputers with $\sim 10$PFLOPS. The validation of our Boltzmann solver has been conducted in a series of papers: the standard tests in static matter distributions meant for radiation transport codes were done in~\citet{2012ApJS..199...17S};~\citet{2014ApJS..214...16N} coupled the Boltzmann solver with a hydrodynamics code of their own construction and tested in dynamical settings the capability of the integrated code in treating special relativistic effects;~\citet{2017ApJS..229...42N}, on the other hand, tested a new module implemented for the tracking of the motion of a proto neutron star (PNS) with a moving grid; very recently~\citet{2017ApJ...847..133R} made a detailed comparison with another Boltzmann solver based on the Monte Carlo method using snapshots from our 2D and 1D simulations and calculating steady-state neutrino distribution functions in the static fluid backgrounds. Having established reliability of our code with these test computations, we now proceed to present the first series of multi-D simulations of CCSNe with the full Boltzmann neutrino transport. In this paper we also pay attention to the neutrino angular distributions in momentum space, which are what the Boltzmann solver is meant for in the first place.
% and, in fact, allow us to validate the approximations commonly used in the literature.
% We pick up the ray-by-ray and M1 methods and compare their prescriptions with the result of our simulations.
 Throughout this paper, Greek and Latin subscripts denote spacetime and spatial components, respectively. We use the metric signature $- + + +$. Unless otherwise stated, we work in units with $c=G=1$, where $c$ is the speed of light and $G$ is the gravitational constant.

%%%%%%%%%%%%%%%%%%%%%%%%%%%%%%%%%%%

\begin{figure*}
\vspace{15mm}
\epsscale{0.9}
\plotone{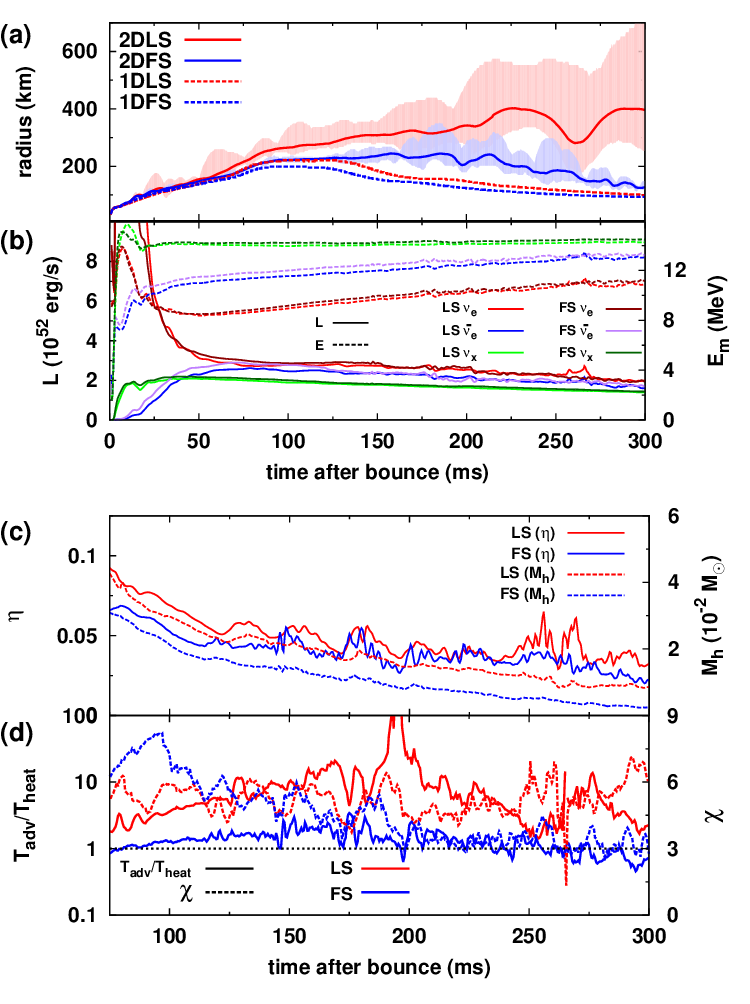}
\caption{(a) Shock radii as functions of time. The color-shaded regions show the ranges of the shock radii, red for the LS EOS and blue for the FS EOS. The solid lines are the angle-average values. For comparison, the corresponding results in spherical symmetry are displayed with dashed lines. (b) Time evolutions of the angle-integrated luminosities ($L$, solid lines) and the angle-averaged mean energies ($E_{\rm m}$, dashed lines) for different species of neutrinos. Both of them are measured at $r=500{\rm km}$. (c) Neutrino-heating efficiency (solid lines) and total mass in the gain region (dashed lines). The heating efficiency is defined as the ratio of the energy deposition rate in the gain region to the sum of the neutrino luminosities of $\nu_{\rm e}$ and $\bar{\nu}_{\rm e}$ (d) The ratio of the advection to heating timescales ($T_{\rm{adv}}/T_{\rm{heat}}$, with solid lines) and the $\chi$ parameter (dashed lines). The dotted black line represents $T_{\rm{adv}}/T_{\rm{heat}}=1$ and $\chi=3$ for reference.
\label{tevo}} 
\end{figure*}

%%%%%%%%%%%%%%%%%%%%%%%%%%%%%%%%%
%%%%%%%%%%%%%%%%%%%%%%%%%%%%%%%%%

\begin{figure*}
\vspace{15mm}
\epsscale{1.0}
\plotone{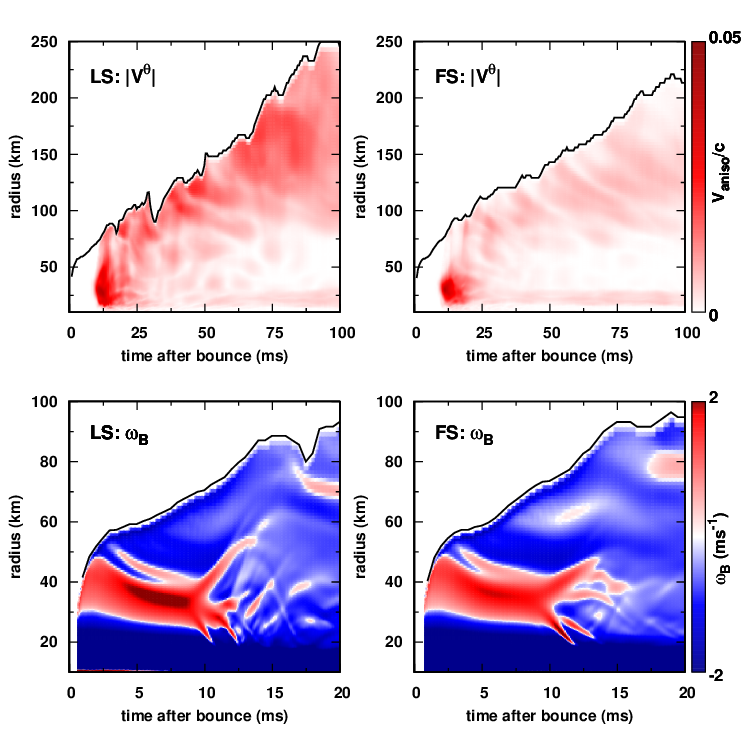}
\caption{Color contours showing time evolutions of the radial profile of angle-averaged lateral velocities ($|v^{\theta}|$) until $100{\rm ms}$ after bounce (top) and of Brunt-V\"{a}is\"{a}l\"{a} frequencies in the very early post-bounce phase up to $20{\rm ms}$ (bottom). Left and right panels present LS- and FS EOS models, respectively. The solid line indicates the minimum shock radius in each panel. Note that a positive (negative) sign is assigned to imaginary (real) Brunt-V\"{a}is\"{a}l\"{a} frequencies in this figure for convenience.
\label{convandvanaiso}} 
\end{figure*}

%%%%%%%%%%%%%%%%%%%%%%%%%%%%%%%%%

\section{Methods and Models} \label{sec:method}

We solve numerically the equations of neutrino-radiation hydrodynamics. 
We apply the so-called discrete-ordinate (DO) method to the Boltzmann equations for neutrino transport, taking fully into account special relativistic effects by virtue of a two-energy-grid technique~\citep{2014ApJS..214...16N}. It has already incorporated general relativistic capabilities as well, a part of which is utilized to track the proper motion of PNS~\citep{2017ApJS..229...42N}.
 The hydrodynamics and self-gravity are still Newtonian: the so-called central scheme of second-order 
accuracy in both space and time is employed for the former and the Poisson equation is solved for the latter.

It should be noted that our treatment of neutrino transport is essentially different from other approximate methods such as the M1 scheme that are commonly employed in the currently most elaborate supernova simulations and are based on the truncated moment formalism one way or another. It is combined with the ray-by-ray approximation in some applications (see e.g., \citet{2012ApJ...756...84M}). In the moment formalism the Boltzmann equation is angle-integrated in momentum space to obtain an infinite number of equations for angular moments, which are then truncated at some order somehow (see Sec.~\ref{sec:distmom} for more details). Such approximations reduce the computational cost drastically. On the other hand, they inevitably introduce the so-called closure relation among lo-order moments, which are the artificial prescriptions to make the truncated equations self-contained. Although the validity of those prescriptions has been assessed for spherically symmetric cases in the literature \citep{2017ApJ...847..133R}, it remains to be demonstrated in multi-dimensional and, more importantly, dynamical settings. In sharp contrast, our approach does not employ any such artificial prescription in the neutrino transport except for the finite-discretization of the Boltzmann equation, which is all but mandatory for this sort of simulations.

We adopt spherical coordinates $(r, \theta)$ covering $0 \le r \le 5000{\rm km}$ and $0^{\circ} \le \theta \le 180^{\circ}$ in the meridian section. 
We deploy $384(r) \times 128(\theta)$ grid points. Momentum space is also discretized non-uniformally with $20$ energy mesh points covering $0 \le \varepsilon 
\le 300{\rm MeV}$ and $10(\bar{\theta}) \times 6(\bar{\phi})$ angular grid points over the entire solid angle. The polar and azimuthal angles $(\bar{\theta}, \bar{\phi})$ 
are locally measured from the radial direction. Three neutrino species are distinguished: electron-type neutrinos $\nu_{\rm e}$, electron-type anti-neutrinos 
$\bar{\nu}_{\rm e}$ and all the others collectively denoted by $\nu_x$.

We pick up a non-rotating progenitor model of 11.2~$M_{\odot}$ from~\citet{2002RvMP...74.1015W}.
We employ two nuclear EOS's: Lattimer \& Swesty's EOS with the incompressibility of $K=220{\rm MeV}$~\citep{1991NuPhA.535..331L} and Furusawa's EOS derived from H. Shen's relativistic mean-field EOS with the TM1 parameter set~\citep{2011ApJ...738..178F,2013ApJ...772...95F}; the former is softer than the latter (see \citet{2004NuPhA.730..227S}). In the following, they are referred to as the "LS" and "FS" EOS's, respectively\footnote{The maximum gravitational masses at zero temperature and non-rotating neutron stars are $2.02 M_{\odot}$ for LS EOS and $2.21 M_{\odot}$ for FS EOS, respectively.}. The choice of EOS's is simply based on the fact that most of previous simulations employed one of these EOS's. We are currently running similar simulations, but with another EOS: Togashi's nuclear EOS based on the variational method with realistic nuclear potentials \citep{2013NuPhA.902...53T} extended by \citet{2017PhRvC..95b5809F} to sub-nuclear densities; it takes into account the full ensemble of heavy nuclei in nuclear-statistical equilibrium (NSE). The results will be reported elsewhere \citep{Nagakuraprep}.
Neutrino-matter interactions are based on those given by~\citet{1985ApJS...58..771B} but we have implemented
the up-to-date electron capture rates for heavy nuclei~\citep{2010NuPhA.848..454J,2000NuPhA.673..481L,2003PhRvL..90x1102L}; they are calculated based on the abundance of heavy nuclei obtained in the FS EOS; the same rates are employed in the LS EOS model just for simplicity; note also that the LS EOS employs a single-nucleus approximation and the detailed information on the population of various nuclei is unavailable. In the current simulations we incorporated the non-isoenergetic scatterings on electrons and positrons as well as the bremsstrahlung in nucleon collisions. We refer readers to~\citet{2014ApJS..214...16N,2017ApJS..229...42N,2012ApJS..199...17S} for more details of our code.

We start the simulations in spherical symmetry and switch them to axisymmetric computations at $\sim1{\rm ms}$ after core bounce when a negative entropy gradient starts to develop behind the shock wave.
We seed by hand at this point of time perturbations of $0.1\%$ in the radial velocities at $ 30 \le  r \le  50{\rm km}$, where convection is expected to occur (see Fig.~\ref{convandvanaiso}). Note that we do not explicitly consider possible turbulent motions that have already existed in the progenitors before collapse. We then expect in non-rotating models that non-radial motions develop initially in the convectively unstable region, and then spread in the rest of the post-shock flow.
% Note that in our simulations we ignore possible effects of pre-collapse convections in the progenitor (see e.g., \citet{2017arXiv170500620M})}.
Each model is run up to $t=300{\rm ms}$ after bounce.

%%%%%%%%%%%%%%%%%%%%%%%%%%%%%%%%%

\begin{figure}
\vspace{15mm}
\epsscale{1.1}
\plotone{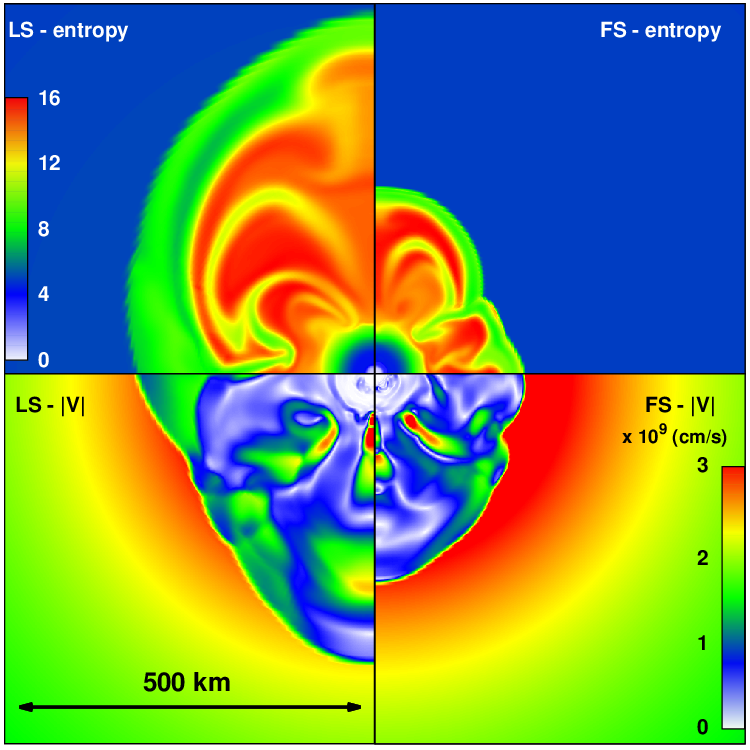}
\caption{Snapshots of entropy per baryon (upper) and fluid-speed (lower) at $t=200{\rm ms}$. Left and right panels are for the LS- and FS EOS, respectively.
   \label{2Dsnap200ms}} 
\end{figure}

%%%%%%%%%%%%%%%%%%%%%%%%%%%%%%%%%

%

%The neutrino-driven convection is dominant over the standing accretion shock instability (SASI) which conforms to the $\chi$ parameter~\citep{2006ApJ...652.1436F,2014ApJ...786..118I} being consistently larger than 3 during the entire post-bounce phase (Fig.~\ref{tevo}(c)).

%Note that {\bf the post-shock morphology} is globally asymmetric as shown in Fig.~\ref{2Dsnap200ms}. 

% with the $\chi$ parameter being higher than 3 until $\sim250{\rm ms}$

\section{Dynamics} \label{sec:dynamics}
As displayed in Fig.~\ref{tevo}(a), the shock wave produced at core bounce expands rather gradually with time for the LS EOS and its maximum radius reaches $\sim700{\rm km}$ at $t=300{\rm ms}$. For the FS EOS, on the other hand, the shock wave stalls at $r\sim 200{\rm km}$ at $t\sim100{\rm ms}$ and then starts to recede at $t\sim250{\rm ms}$ and shrinks back to $r\sim100{\rm km}$ by $t\sim300{\rm ms}$. Although the time evolutions of the average shock radii of the two models are quite similar to each other until $\sim 60{\rm ms}$ after bounce and their deviations become remarkable thereafter, some differences have already appeared in the post-shock flows by this time in fact.

 In the top two panels of Fig.~\ref{convandvanaiso}, we compare the angle-averaged amplitudes of lateral velocity for the two models. The more reddish the color is, the stronger the lateral motions are. It is apparent that they become appreciable initially at $t \sim 10{\rm ms}$ in a region at $r \sim 30{\rm km}$ almost simultaneously, which marks the onset of the prompt convection. Although the turbulent region extends upwards in both models, the amplitudes of the lateral velocity are larger for the LS EOS than for the FS EOS, indicating that the prompt convection is more vigorous in the former. This trend persists until much later times, though, as is also evident from the figure.

The difference in the strength of the prompt convection may be understood from the difference in the Brunt-V\"{a}is\"{a}l\"{a} frequencies, which are compared in the lower panels of Fig.~\ref{convandvanaiso}. Reddish colors again imply more rapid (exponential) growths of the convection. As expected, the unstable region emerges at $r \sim 30-50{\rm km}$ immediately after the switch to the 2D computations for both models. Although this strongly unstable region persists until $t\sim 15{\rm ms}$ at about the same location for both models, the maximum frequency is larger for the LS EOS. This difference can be traced back to the difference in photodissociations of heavy nuclei by shock heating. In fact, they are stronger in the LS EOS and, as a result, the shock is weakened more severely, producing steeper negative entropy gradients in this case. The initial fluctuations produced this way are carried upwards by acoustic waves, which are also stronger in the LS EOS. As a consequence, the turbulent motions are more vigorous for the LS EOS than for the FS EOS as already mentioned, the fact that has an important implication for later evolutions of the shock waves.

 It is interesting that the neutrino luminosities ($L$) and mean energies ($E_{m}$, defined as the ratio of energy density to number density) are almost identical between the two cases (Fig.~\ref{tevo}(b)). It should be noted, however, that the neutrino-heating efficiency is different, being higher for the LS EOS (see solid lines in Fig.~\ref{tevo}(c)). This is mainly because the total baryon mass in the gain region, where heating dominates over cooling and the net heating occurs, is consistently larger for the LS EOS than for the FS EOS (dashed lines in the same panel). This in turn seems to be a consequence of the turbulent motions that are more vigorous for the LS EOS as we mentioned in the previous paragraphs.

 Figure~\ref{2Dsnap200ms} compares the entropy and velocity distributions between the two models at $t=200 {\rm ms}$. Their post-shock morphologies are quite similar to each other and only the scales are different. In fact, the convection is dominant over SASI in most of the post-bounce phase for both models (see the $\chi$ parameter~\citep{2006ApJ...652.1436F,2014ApJ...786..118I} in Fig.~\ref{tevo}(d)). In the same panel, we also show the ratio of the advection timescale ($T_{\rm adv} = M_{\rm g}/\dot{M}$ with $M_{\rm g}$ and $\dot{M}$ denoting the mass in the gain region and the mass accretion rate, respectively) to the heating timescale ($T_{\rm heat} = |E_{\rm tot}|/\dot{Q}_{\nu}$ with $E_{\rm tot}$ and $\dot{Q}_{\nu}$ being the total energy and the heating rate in the gain region, respectively) as solid lines. One can see that it is consistently larger for the LS EOS than for the FS EOS, meaning that the former has more favorable conditions for shock revival than the latter.

The decline of this ratio near the end of the simulation for the LS EOS in spite of a continuous growth of the maximum shock radius is an artifact originated from our choice of the minimum shock radius in the evaluation of the ratio. As displayed in Fig.~\ref{tevo}(a), the minimum shock radius is still decreasing with time at the end of the simulation. Then the volume of gain region is underestimated and, as a result, $T_{\rm heat}$ is overestimated. The fact that the ratio occasionally exceeds unity but still yields no shock revival for the FS EOS indicates that the criterion is not a rigorous condition, which is understood also from the uncertainty in its definition just mentioned. We do not intend to discuss the applicability of the diagnostics any further in this paper but we still think it is useful in judging, albeit roughly, which model is closer to shock revival.

% Note that the differences in EOS and other microphysics tend to be more remarkable in multi-D than in spherical symmetry~\citep{2016arXiv161105859B,2012ApJ...755..138H}. These results may be subject to change by the inclusion of general relativistic gravity (see~\citet{2012ApJ...756...84M,2013ApJ...765...29C,2015arXiv151107443O}), which is still lacking in the current simulations and should be studied further.

%%%%%%%%%%%%%%%%%%%%%%%%%%%%%%%%%

\begin{figure}
\vspace{15mm}
%\epsscale{1.2}
\epsscale{1.2}
%\plotone{test_prlv2_nue.eps}
\plotone{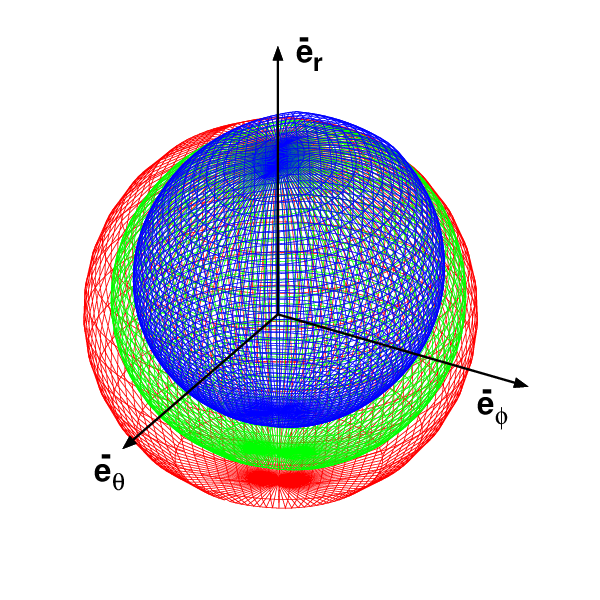}
\caption{Angular distributions of $\nu_{\rm e}$ in momentum space at 
 $t_{\rm pb} = 15{\rm ms}$ for the LS EOS. Different colors correspond to different radial positions (red: $r=23{\rm km}$, green: $r=39{\rm km}$, 
 blue: $r=49{\rm km}$) along the radial ray with the zenith angle of $\theta = 8 \pi/15$. The neutrino energy is $\varepsilon = 11.1{\rm MeV}$ in the fluid-rest frame.
\label{wired}} 
\end{figure}

\begin{figure*}
\vspace{15mm}
\epsscale{1.2}
\plotone{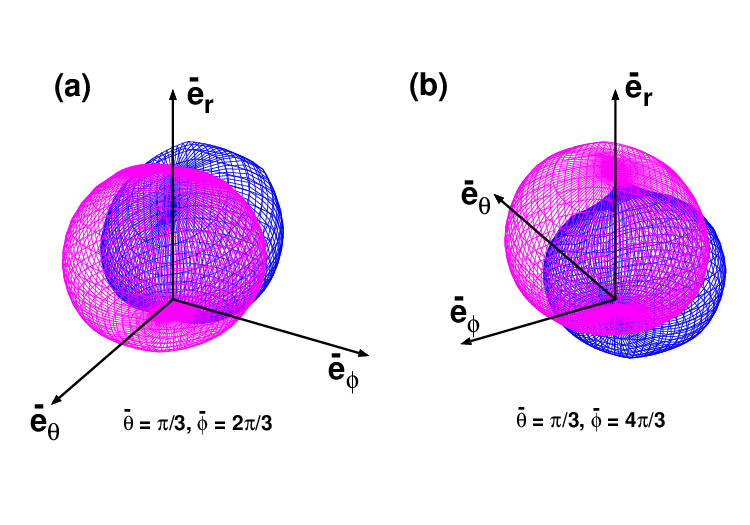}
\caption{Similar to Fig.~\ref{wired} but the deviations from spherical symmetry emphasized and viewed from different angles: (a) $\bar{\theta} = \pi/3$ and $\bar{\phi} = 2 \pi/3 $ (b) $\bar{\theta} = \pi/3$ and $\bar{\phi} = 4 \pi/3 $. In each panel, the minimum is subtracted isotropically from the original angular distribution and the resultant distribution is normalized so that the maximum value should be always identical. The blue surface corresponds to the one with the same color in Fig.~\ref{wired} while the purple surface shows another subtracted surface at the same radius but at a different zenith angle, $\theta =  17\pi/45$.
\label{wired_v2}} 
\end{figure*}

%%%%%%%%%%%%%%%%%%%%%%%%%%%%%%%%%

\section{$\nu$-Distributions in Momentum Space} \label{sec:distmom}

Next we turn our attention to novel features of the neutrino distributions in momentum space. We find in our calculations significant non-axisymmetry with respect to the radial direction in the neutrino angular distributions. It is produced by lateral inhomogeneities in matter, which are in turn generated by hydrodynamical instabilities. The asymmetry hence appears inevitably in multi-D simulations.

Figure~\ref{wired} shows as an example the angular distributions of $\nu_{\rm e}$ with an energy of $\varepsilon = 11.1{\rm MeV}$ at three different radial positions. Each surface displays the neutrino distribution function for different propagation directions normalized by the maximum value in the fluid-rest frame. Colors of the surfaces denote the locations on an arbitrarily chosen radial ray. The angular distribution is almost isotropic at $r=23{\rm km}$ (red surface) while they become forward peaked (green and blue surfaces) as the radius increases, a fact that is well known. What is really new here is that they are non-axisymmetric with respect to the radial direction, which is more apparent in Fig.~\ref{wired_v2}, in which the isotropic contributions are subtracted from the original distributions and the resultant ones are re-normalized by their maximum values. Note that the feature is robust, occurring irrespective of neutrino energies or species.

It should be mentioned, however, that the non-axisymmetric angular distributions obtained in the current simulations still have a symmetry with respect to the azimuthal angle ($\bar{\phi}$) in momentum space. This is due to the fact that these are non-rotating models and there is a mirror symmetry with respect to the plane spanned by $\mbox{\boldmath $\bar{e}$}_r$ and $\mbox{\boldmath $\bar{e}$}_{\theta}$ in momentum space in the absence of rotation. Once rotation is taken into account, the symmetry is lost even in (spatial) axisymmetry. This is the reason why we do not assume this symmetry in our code. In 3D simulations, no symmetry remains in the angular distribution in momentum space. Its characterization is an interesting subject of spatially 3D supernova simulations with multi-angle neutrino transport, which are currently being undertaken and will be reported elsewhere later.

%%%%%%%%%%%%%%%%%%%%%%%%%%%%%%%%%

\begin{figure*}
\vspace{15mm}
\epsscale{0.8}
\plotone{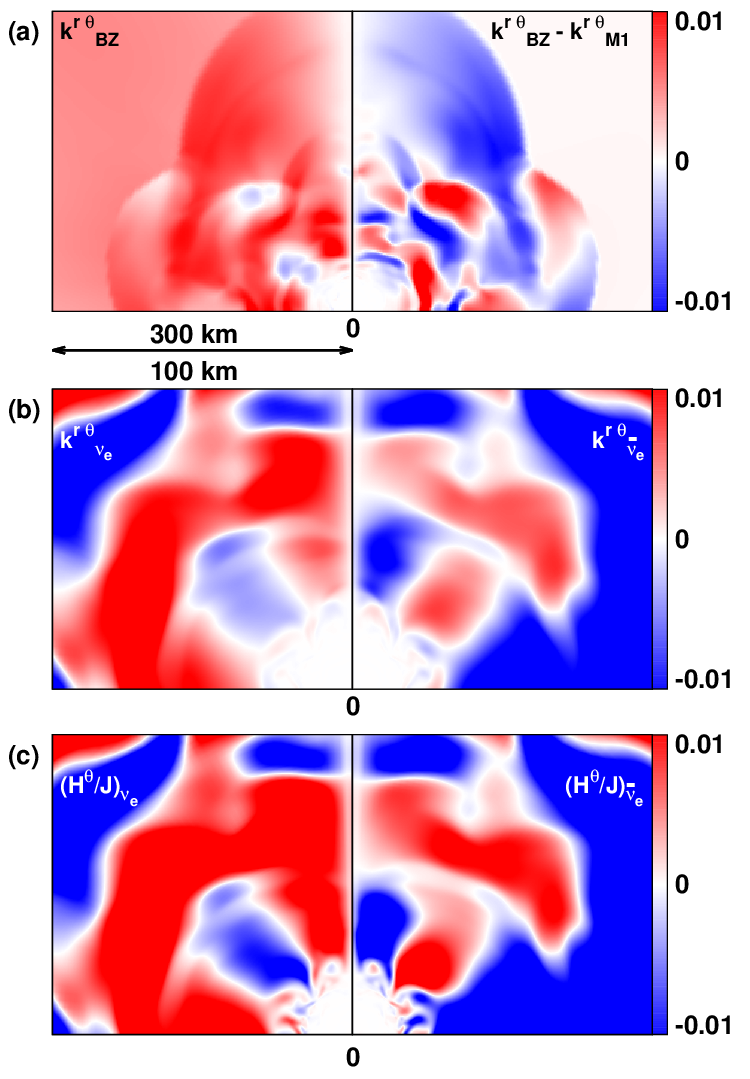}
\caption{(a) The $(r {\theta})$ component of the Eddington tensor ($k^{r \theta}$) for $\nu_{\rm e}$ in the northern hemisphere obtained in our simulation for the FS EOS (left) and its deviation from the M1 prescription (right). The values of $k^{r \theta}$ are evaluated at the mean neutrino energy at each point. (b) $k^{r \theta}$ for $\nu_{\rm e}$ (left) and $\bar{\nu}_{\rm e}$ (right) on a smaller spatial scale of $100{\rm km}$. The neutrino energy is fixed to $8.53{\rm MeV}$ in the fluid-rest frame. (c) Same as the panel (b) but for $H^{\theta}/J$ with \mbox{\boldmath $H$} and $J$ being the energy-flux and energy densities measured in the fluid-rest frame, respectively. The time is $t=190{\rm ms}$ in all cases.
\label{Edten}} 
\end{figure*}

%%%%%%%%%%%%%%%%%%%%%%%%%%%%%%%%%

The multi-angle treatment of neutrino transport in our simulations enables us to evaluate the so-called Eddington tensor ($k^{ij}$), which characterizes these non-axisymmetric angular distributions more quantitatively.
% Note that hereinafter Latin subscripts denote the spatial components of tensors alone while Greek letters are used for both spatial and temporal components.
 The Eddington tensor is obtained from the neutrino distribution function ($f$) as follows: we first define the second angular moment $M^{\mu \nu}$ as
\begin{eqnarray}
M^{\mu \nu}(\varepsilon) \equiv \frac{1}{\varepsilon} \int f(\varepsilon,\Omega_m) p^{\mu}  p^{\nu} d\Omega_m,
\label{eq:Moment}
\end{eqnarray}
where $p^{\mu}$ is the four-momentum of neutrino and $\varepsilon$ and $\Omega_m$ are the corresponding energy and solid angle measured in the fluid-rest frame; then the Eddington tensor $k^{ij}$ is given as
\begin{eqnarray}
k^{ij}(\varepsilon) \equiv \frac{P^{ij}(\varepsilon)}{E(\varepsilon)}, \label{eq:defkij}
\end{eqnarray}
where $P^{ij}$ and $E$ are defined from $M^{\mu \nu}$ as
\begin{eqnarray}
&&P^{ij}(\varepsilon) \equiv \gamma_{\hspace{1mm} \mu}^{i} \gamma_{ \hspace{1mm}  \nu}^{j} M^{\mu \nu} (\varepsilon) \label{eq:Pij}, \\
&&E(\varepsilon) \equiv n_{\mu} n_{\nu} M^{\mu \nu}(\varepsilon) \label{eq:Emom},
\end{eqnarray}
with $n_{\mu}$ and $\gamma_{\hspace{1mm} \mu}^{i}$ (= $\delta_{\hspace{1mm} \mu}^{i} + n^{i} n_{\mu})$ being the unit vector orthogonal to a hypersurface of constant coordinate time and the projection tensor onto this hypersurface, respectively.

We pay particular attention here to one of the off-diagonal components of the Eddington tensor, $k^{r \theta}$, which are zero in spherical symmetry in space, i.e., they are a measure of genuinely multi-dimensional transfer.
% {\bf Note that the ray-by-ray(-plus) approximation neglects $k^{r \theta}$ completely while two-moment method uses rather makeshift prescriptions to compute it.}
%~\citep{2012ApJ...756...84M,2013ApJ...767L...6B,2014ApJ...786...83T,2016ApJ...818..123B,2015ApJ...807L..31L,2015ApJ...801L..24M,2016ApJ...825....6S,2017MNRAS.468.2032A}
%Note that the existence of non-vanishing $k^{r \theta}$ may be an important warning sign that the ray-by-ray(-plus) approximation~\citep{2012ApJ...756...84M,2013ApJ...767L...6B,2014ApJ...786...83T,2016ApJ...818..123B,2015ApJ...807L..31L,2015ApJ...801L..24M,2016ApJ...825....6S,2017MNRAS.468.2032A}, which neglects $k^{r \theta}$ completely, may yield inaccurate results.
The left panel in Fig.~\ref{Edten}(a) shows $k^{r \theta}$ for $\nu_{\rm e}$ with the mean energy at each point. As expected, it is almost zero inside the PNS, where matter is opaque enough to make the neutrino distribution isotropic. It becomes non-zero outside the PNS, however, and increases with radius in accord with the appearance of the non-axisymmetric structures in the neutrino angular distribution (see Fig.~\ref{wired}). In fact, the $k^{r \theta}$ corresponds to the mode with $\ell=2, m=1$ in the spherical harmonics expansion of the distribution function.

% The right panel in Fig.~\ref{Edten}(a) shows, on the other hand, {\bf the difference between our results and prescriptions by} the two-moment approximation.
 The right panel in Fig.~\ref{Edten}(a) compares $k^{r \theta}$ obtained from our simulation with that which is evaluated according to the M1 prescription: the Eddington tensor in the M1 prescription ($k^{ij}_{\rm M1}$) is obtained by replacing $P^{ij}$ in Eq.~(\ref{eq:Pij}) with
\begin{eqnarray}
P^{ij}_{\rm M1}(\varepsilon) = \frac{3 \zeta(\varepsilon) - 1}{2} P^{ij}_{\rm thin}(\varepsilon)  
                             + \frac{3(1 - \zeta(\varepsilon))}{2} P^{ij}_{\rm thick}(\varepsilon) ,
\label{eq:PijM1}
\end{eqnarray}
where $\zeta$ is referred to as the variable Eddington factor, which we set as
\begin{eqnarray}
\zeta(\varepsilon) = \frac{3 + 4\bar{F}(\varepsilon)^2}{5+2\sqrt{4-3\bar{F}(\varepsilon)^2}}.
\label{eq:vef}
\end{eqnarray}
In this expression, $\bar{F}$ denotes the so-called flux factor, which is the energy-flux normalized with the energy density in the fluid-rest frame. The flux factor that we use in this paper is measured in the fluid-rest frame (see \citet{2011PThPh.125.1255S} for another option);
\begin{eqnarray}
\bar{F}(\varepsilon) =  \Biggl(  \frac{h_{\mu \nu} H^{\mu}(\varepsilon) H^{\nu}(\varepsilon) } {J(\varepsilon)^2}
 \Biggr)^{1/2},
\label{eq:fluxfactor}
\end{eqnarray}
where $J$ and $H^{\mu}$ can be expressed in terms of $M^{\mu \nu}$ as
\begin{eqnarray}
&&J(\varepsilon) = u_{\mu} u_{\nu} M^{\mu \nu} (\varepsilon), \nonumber \\
&&H^{\mu}(\varepsilon) = - h_{\hspace{1mm} \alpha}^{\mu} u_{\beta} M^{\alpha \beta} (\varepsilon),
\label{eq:defHij}
\end{eqnarray}
with $u^{\mu}$ and  $h_{\hspace{1mm} \nu}^{\mu} (= \delta_{\hspace{1mm} \nu}^{\mu} + u^{\mu} u_{\nu})$ being the fluid four velocity and the projection tensor onto the fluid-rest frame, respectively. The optically thick and thin limits of $P^{ij}$ are denoted by $P^{ij}_{\rm thick}$ and $P^{ij}_{\rm thin}$~\citep{2015MNRAS.453.3386J,2011PThPh.125.1255S,2015arXiv151107443O,2016ApJS..222...20K}, which are written as
\begin{eqnarray}
&&P^{ij}_{\rm thick}(\varepsilon) = J(\varepsilon) \frac{\gamma^{ij} + 4 V^{i} V^{j} }{3} + H^{i}(\varepsilon) V^{j} + V^{i} H^{j}(\varepsilon)  , \nonumber \\
&&P^{ij}_{\rm thin}(\varepsilon) = E(\varepsilon) \frac{F^{i}(\varepsilon) F^{j}(\varepsilon)}{F(\varepsilon)^2},
\label{eq:Pijthickthin}
\end{eqnarray}
where $V^{i}$ denotes the three dimensional vector of fluid velocity. $F^{i}$ can be expressed in terms of $M^{\mu \nu}$ as
\begin{eqnarray}
F^{i}(\varepsilon) = - \gamma_{\hspace{1mm} \mu}^{i} n_{\nu}  M^{\mu \nu} (\varepsilon).
\label{eq:defFi}
\end{eqnarray}
As clearly seen in this panel, the values of $k^{r \theta}$ are substantially different between the two cases. We find that such discrepancies in $k^{r \theta}$ are rather generic, being insensitive to the choice of the prescription for the Eddington factor (see~\citet{2015MNRAS.453.3386J} for various options). They are also systematic in the sense that the increase in the number of grid points in the M1 prescription does not reduce the difference. This is in contrast to our approach, in which the accuracy is simply improved with the resolution.

%while the rather low angular resolution in momentum space (see \citet{2017ApJ...847..133R} for more details) is admittedly a problem of our current simulations but

Moreover, we find in $k^{r \theta}$ an intriguing correlation/anti-correlation between $\nu_{\rm e}$ and $\bar{\nu}_{\rm e}$. The two panels of Fig.~\ref{Edten}(b) compare $k^{r \theta}$ for $\nu_{\rm e}$ and $\bar{\nu}_{\rm e}$ with the same energy of $\varepsilon = 8.5{\rm MeV}$. As can be seen in these panels, they are anti-correlated with each other in the vicinity of PNS ($\lesssim 50{\rm km}$) whereas they are positively correlated at larger radii ($> 80{\rm km}$). The anti-correlation is particularly remarkable for low-energy neutrinos with $\lesssim 10{\rm MeV}$. We find that the sign of $k^{r \theta}$ roughly coincides with that of the lateral neutrino flux, which is shown in Fig.~\ref{Edten}(c). In fact, it is apparent that the lateral flux is oriented in the opposite directions for $\nu_{\rm e}$ and $\bar{\nu}_{\rm e}$. This is in turn due to the Fermi-degeneracy of $\nu_{\rm e}$ at $r \lesssim 30{\rm km}$, which produces opposite trends in the number densities of $\nu_{\rm e}$ and $\bar{\nu}_{\rm e}$. Since neutrinos flow from high to low $\nu$ number density regions in the diffusion regime, the fluxes of $\nu_{\rm e}$ and $\bar{\nu}_{\rm e}$ should be naturally anti-correlated as a result of the opposite trend in the number densities of $\nu_{\rm e}$ and $\bar{\nu}_{\rm e}$. We do not know for the moment how this anti-correlation in the fluxes is transferred to that in $k^{r \theta}$. It will be necessary to analyze more in detail the equations of motion for higher moments including $k^{r \theta}$.

 Importantly, the anti-correlation is then carried to larger radii by the radial flux and remains non-vanishing even at $r \sim 50{\rm km}$, where $\nu_{\rm e}$ is no longer degenerate. On the other hand, at even larger radii, where matter is optically thin to neutrinos, $k^{r \theta}$ is correlated with the local lateral velocity of matter due to relativistic aberration. Note that this positive correlation at large distances is less remarkable than the anti-correlation in the vicinity of PNS (see the equatorial region in Fig.~\ref{Edten}(b)), since the angular distribution is no longer determined locally and the correlation is somewhat smeared out.

As will be discussed in Sec.~\ref{sec:importance}, the appropriate treatment of $k^{r \theta}$ is related with the accurate calculation of the neutrino flux, in particular its lateral component (see Eqs.~(\ref{eq:Frasymevo}) and (\ref{eq:Fthasymevo})). It is true that these correlation/anti-correlation look rather minor but they may play an important role through the lateral fluxes of neutrinos. In fact they clearly indicate the intricacy of neutrino transport in non-spherically dynamical settings. It will be interesting to see how well the M1 scheme can reproduce these features and to conceive possible improvements of its prescription.

\section{Angular Resolution in momentum space} \label{sec:resolution}

%%%%%%%%%%%%%%%%%%%%%%%%%%%%%%%%%
\begin{figure*}
\vspace{15mm}
\epsscale{1.2}
\plotone{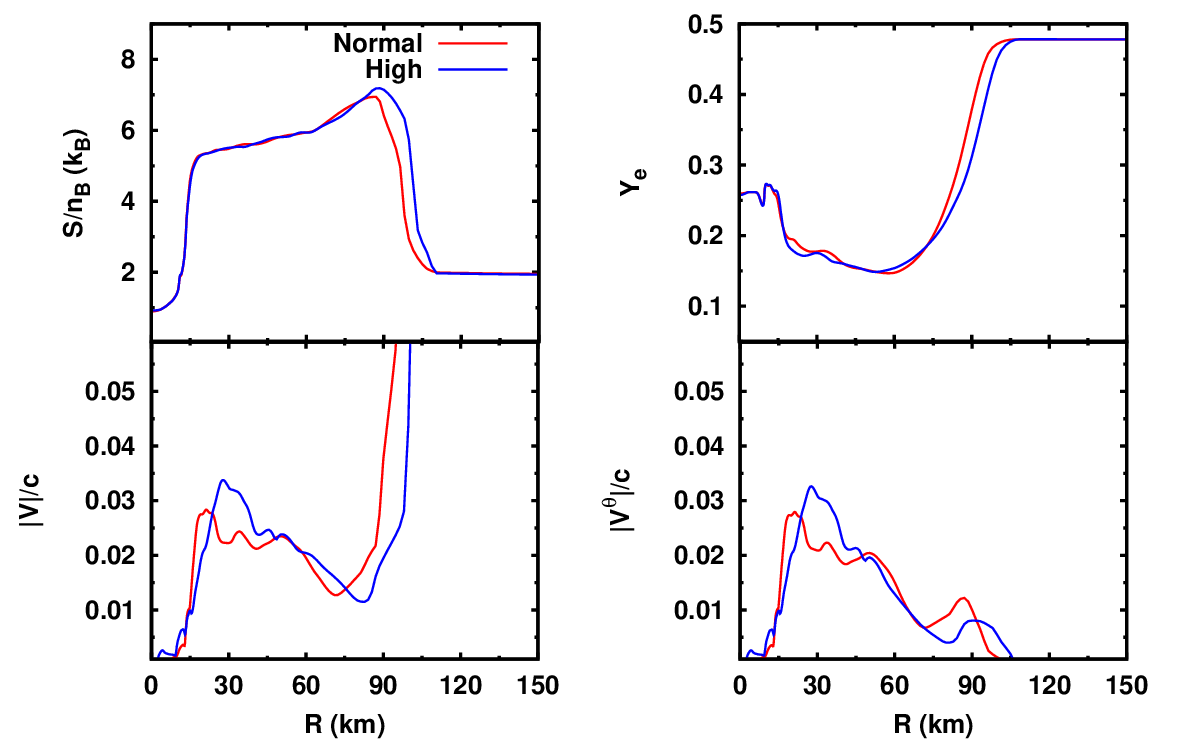}
\caption{Angle-averaged radial profiles of fluid quantities. Upper left: entropy per baryon. Upper right: electron fraction. Bottom left: fluid speed. Bottom right: absolute values of lateral velocity. The red line shows the result of the normal-resolution while the blue lines correspond to the high-resolution simulation. The time is $t=15$ms post-bounce.
\label{hydro_reso}} 
\end{figure*}
%%%%%%%%%%%%%%%%%%%%%%%%%%%%%%%%%
%%%%%%%%%%%%%%%%%%%%%%%%%%%%%%%%%
\begin{figure*}
\vspace{15mm}
\epsscale{1.2}
\plotone{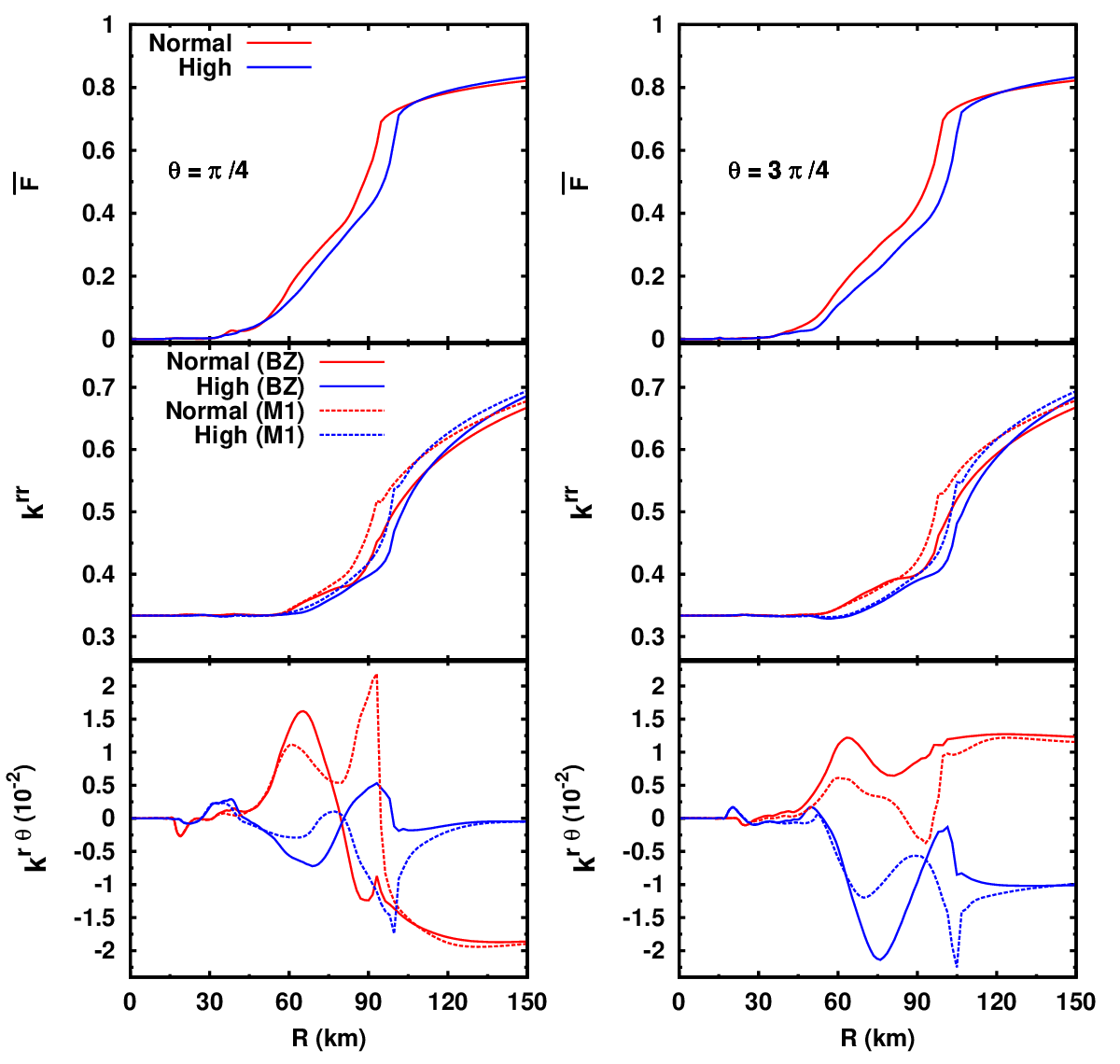}
\caption{The flux factor (top), and the $rr$ (middle) and $r \theta$ (bottom) components of the Eddington tensor for electron-type neutrinos. The left column presents the radial profiles along the radial ray with $\theta = \pi/4$, while the right one displays the same quantities but for $\theta = 3 \pi/4$. The colors of lines and the time of the snap shot ($t=15$ms post-bounce) are the same as in Fig.~\ref{hydro_reso}.
\label{neutrino_reso}} 
\end{figure*}
%%%%%%%%%%%%%%%%%%%%%%%%%%%%%%%%%

This study is the first ever attempt to perform spatially 2D supernova simulations with multi-angle and multi-energy neutrino transport, taking into account all special relativistic effects completely. It is a legitimate concern, however, that the current simulations may not have a sufficient numerical resolution especially in momentum space \citep{2017ApJ...847..133R}. In this section we hence discuss this resolution issue, focusing on the angular resolution in momentum space.

 For that purpose we perform a new high-resolution simulation for the early post-bounce phase, whereas for the discussion of the late post-bounce phase we employ the results of our previous analyses \citep{2017ApJ...847..133R} of time-independent solutions of the Boltzmann equations for neutrinos in given matter distributions; close comparisons were made with the data obtained with Monte Carlo Simulations \citep{2015ApJ...813...38R}. Note that although the use of the time-independent solutions for the fixed matter distributions enabled us to conduct rigorous comparisons, its applicability may be limited to the late post-bounce phase, where the time scale of variations in the background is indeed long. For the earlier phase, however, we need to consider time-dependent solutions. We hence run a higher-resolution simulation, in which the time evolutions of both neutrino and matter distributions are computed for only $15$ms from the bounce with the LS EOS. We compare the results so obtained with the original ones to see to what extent the angular resolution could affect the outcome. Note, however, that the comparisons are not so clear-cut as in the previous paper, since the matter dynamics in this phase is chaotic and small perturbations induced by the change in the angular resolution modify not only the neutrino distributions but also the matter configurations in the background substantially.

\citet{2017ApJ...847..133R} demonstrated that our Boltzmann solver tends to underestimate the forward peak in the angular distributions of neutrinos in momentum space at large radii if the number of the angular mesh points is not large enough. This is actually just as expected and was indeed pointed out by \citet{1999A&A...344..533Y} in their 1D study. As a matter of fact, neutrinos are moving almost radially at large distances from the neutrino sphere no matter what happens to them at small radii and if the angular spread becomes smaller than the smallest width of the angular bin employed in the Boltzmann solver, it is no longer resolved.

Such properties of our Boltzmann solver should have some implications for the success or failure of explosion in our simulations, since the underestimation of the forward peak in the angular distribution in momentum space leads in turn to the overestimation of the local number density of neutrinos and, as a result, the overestimation of neutrino heating in the gain region. On the other hand, \citet{2017ApJ...847..133R} also found that the finite energy resolution tends to underestimate the neutrino heating. We then surmise from these results that the volume-integrated net energy deposition in the gain region is probably underestimated in the current simulations by a few percent.

% lower than one for the highest resolution in both 2D and 1D cases, meaning that our current resolution probably underestimates several percent error for neutrino heating, which would artificially suppress for the development of explosion. 

%Note that our energy resolution is comparable or even higher than other CCSNe simulations. Since weak reaction rate sensitively depends on the neutrino energy, we 

For the study of the resolution dependence in the early post-bounce phase, we conduct a high-resolution simulation for a short period as mentioned earlier. This time the matter distribution is not fixed but calculated just as in the ordinary run. We deploy $14(\bar{\theta}) \times 10(\bar{\phi})$ angular grid points over the entire solid angle while space and energy grids are unchanged from the normal run. In Fig.~\ref{hydro_reso}, we compare the radial profiles of some angle-averaged quantities at $15$ms after bounce between the models with the normal and high angular resolutions. As can be seen in this figure, the prompt shock wave is a bit faster and reaches a larger radius in the high-resolution model than in the normal-resolution model (upper left panel); in association with this, the deleptonization behind the shock is slightly stronger in the former around $ 20 \le  r \le  40{\rm km}$ (upper right). These are all attributed to the fact that the high-resolution simulation experiences a stronger prompt convection. This is indeed corroborated both in the fluid velocity and their lateral component in the convectively unstable region: they are a little larger in the high-resolution simulation consistently. As mentioned earlier, however, matter motions in this region are stochastic due to the chaotic nature of convection. The results would be different substantially if, for example, the initial time is changed even slightly. It is also difficult to isolate the influence of the angular resolution on the neutrino transport alone. More detailed resolution studies in dynamical settings will be reported elsewhere. With these caveats in mind, we will further compare some quantities of relevance in neutrino transport.

%%%%%%%%%%%%%%%%%%%%%%%%%%%%%%%%%
\begin{figure*}
\vspace{15mm}
\epsscale{1.2}
\plotone{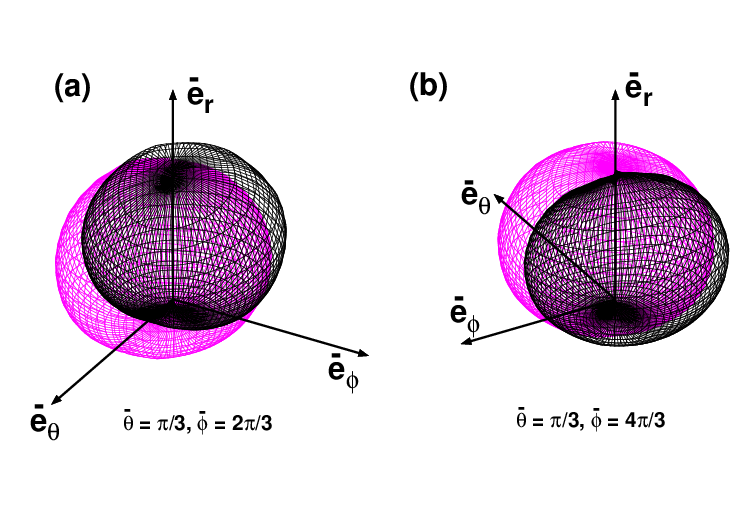}
\caption{The same picture as in Fig.~\ref{wired_v2} but for two different angular resolutions. The purple wired frame is identical to the same purple one in Fig.~\ref{wired_v2}. The black one is a high-resolution counterpart.
\label{wired_reso}} 
\end{figure*}
%%%%%%%%%%%%%%%%%%%%%%%%%%%%%%%%%

Figure~\ref{neutrino_reso} displays the radial profiles along two radial rays with $\theta = \pi/4$ (left column) and $ \theta = 3 \pi/4$ (right column) of some relevant quantities in the $\nu_e$ distribution at $15$ms after bounce. The neutrino energy is set to the average value at each point. In the top panels, the flux factors ($\bar{F}$) defined in Eq.~(\ref{eq:fluxfactor}) are shown. One immediately recognizes that it is systematically smaller for the high-resolution case in the post-shock region. This is not directly related with the angular resolution, though. Instead it is simply because the shock radius is larger in the high-resolution run and, as a result, the flux factor increases more slowly from the optically thick limit ($\bar{F} =0$) to the thin limit ($\bar{F} = 1$). On the other hand, the flux factor is always smaller for the normal case than for the high-resolution case at large radii. This is a direct resolution effect, i.e., the low-resolution simulation fails to reproduce the forward peak in the angular distribution at large radii.

 The rr components of the Eddington tensor, $k^{rr}$, are shown in the middle panels of Fig.~\ref{neutrino_reso}. It is observed that they also increase a bit more slowly initially in the high-resolution run. This is again a mere consequence of the larger shock radius in that case. In these panels, we also display as additional dotted lines the same components of the Eddington tensor that are obtained with the M1 prescription. Except in the inner optically thick region, they are always slightly greater than those obtained with the Boltzmann code for both resolutions. Considering the result in \citet{2017ApJ...847..133R} that low-resolution computations with the Boltzmann solver tend to underestimate $k^{rr}$, one may think that the results of the M1 prescription is closer to the true values. It should be noted, however, that the differences found here in $k^{rr}$ between the Boltzmann and M1 results are larger than those obtained in \citet{2017ApJ...847..133R} (see Fig.17 in their paper). This may imply that the M1 prescription has its own problem in reproducing $k^{rr}$ for highly-time dependent and highly-inhomogeneous matter distributions considered here. This issue will be further studied in our forthcoming paper. It is incidentally pointed out that the M1 prescription needs the flux factor to obtain the Eddington tensor (see Eqs.~(\ref{eq:PijM1}) and (\ref{eq:vef})). In the present comparison it is provided by the Boltzmann solver although it should be calculated on its own in the actual simulations with the M1 approximation. It is hence desirable to make comparisons, employing the results of such M1 simulations, which is another subject worth further investigations.

%Fairly speaking, however, the obtained $k^{rr}$ in M1 prescription uses the flux factor obtained from Boltzmann results, which shows in the upper panels, meaning that M1 prescription still requires to reproduce those of Boltzmann's results if the prescription is correct.

%Since, in terms of spherically symmetric radiation field, the diagonal component are almost identical between Boltzmann and M1 prescription (see Fig.9 in \citet{2017ApJ...847..133R}),

The bottom panels in Fig.~\ref{neutrino_reso} are again the Eddington tensors but for the $r\theta$ component $k^{r \theta}$ this time. It should be noted first that $k^{r \theta}$ is very sensitive to the matter motion in the background. As a result, their profiles are quite different between the normal and high-resolution simulations and it is rather difficult to discuss the convergence in the current dynamical setting. Nevertheless, it is evident that the Boltzmann and M1 results are substantially different from each other even qualitatively in the semi-transparent region although they agree in both the optically thin and thick limits irrespective of resolutions. This is indeed consistent with the findings by \citet{2017ApJ...847..133R}, who also came to the same conclusion that the difference in $k^{r \theta}$ between the Boltzmann transport with multi-angles and the M1 prescription in the semi-transparent regime is intrinsic and never reduced by increasing resolution. As will be demonstrated in Sec.~\ref{sec:importance}, inaccurate $k^{r \theta}$ may give a $\sim 10 \%$ level of errors in the neutrino luminosity and, more importantly, will lead to qualitatively wrong lateral fluxes of neutrinos in the semi-transparent region.

% neutrino flux, which are discussed in Sec.~\ref{sec:importance}.
% This is consistent with our previous study in \citet{2017ApJ...847..133R}) that the difference of $k^{r \theta}$ between multi-angle transports and M1 prescription in the semi-transparent regime is never converged with increasing resolution.

In Fig.~\ref{wired_reso}, we compare the angular distributions in momentum space obtained with the two resolution. Note that, the isotropic contributions are subtracted as previously in these pictures so that the anisotropies could be better recognized. In panel (a), the purple surface is identical to the one presented in Fig.~\ref{wired_v2}, while the black surface is the high-resolution counterpart. In Fig.~\ref{wired_reso}(b), we change the viewing angle to facilitate readers' understanding of the non-axisymmetric features. As mentioned above, since the matter distributions in the background are different between the two cases, the neutrino angular distributions differ qualitatively. It is important, however, that the degree of asymmetry is even more prominent in the high-resolution simulation. This is again consistent with the finding in \citet{2017ApJ...847..133R} that $k^{r \theta}$ tends to be underestimated in low-angular resolution simulations (see the right panel of Fig.15 in their paper).

\section{Possible implications of off-diagonal components on supernova dynamics} \label{sec:importance}

%%%%%%%%%%%%%%%%%%%%%%%%%%%%%%%%%
\begin{figure*}
\vspace{15mm}
\epsscale{1.2}
\plotone{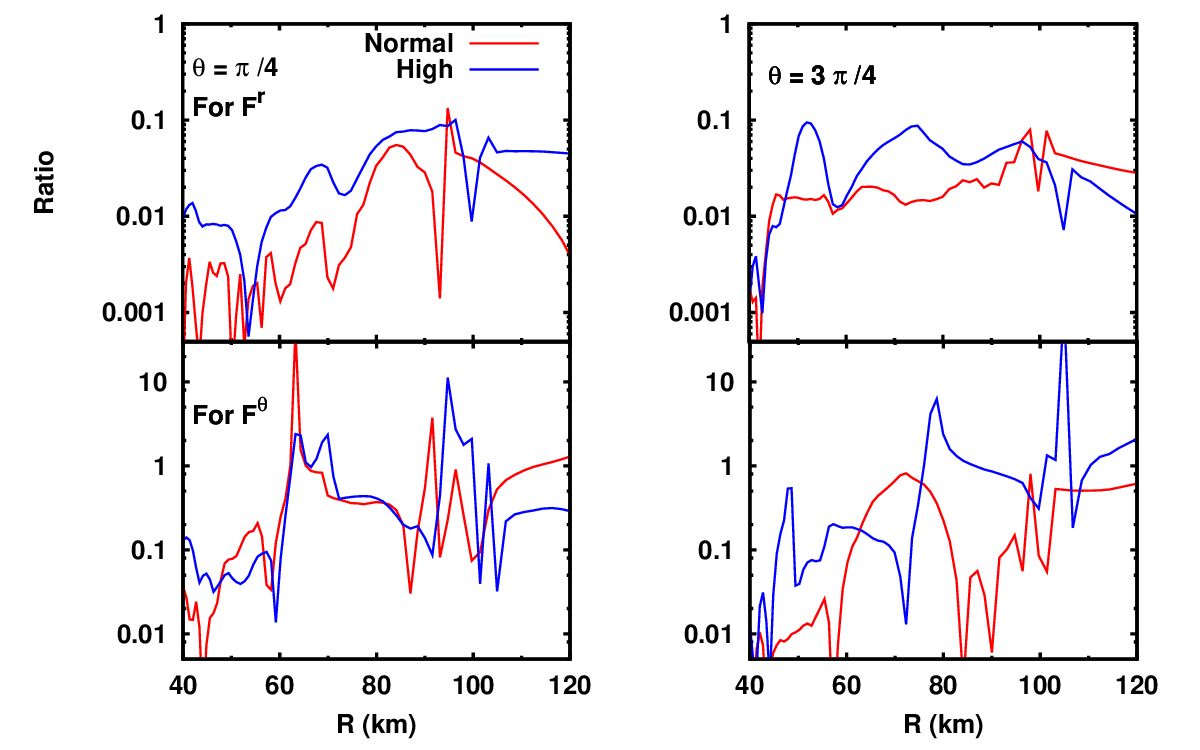}
\caption{The radial profiles of the absolute ratios of $\partial_{\theta} (Ek^{r \theta})/r$ to $\partial_r (E k^{r r})$ (upper panels) and $\partial_r (E k^{r \theta})$ to $\partial_{\theta} (E k^{\theta \theta})/r$ (lower panels). These quantities measure the relative importance of the terms on the right hand side of Eqs.~(\ref{eq:Frasymevo}) and (\ref{eq:Fthasymevo}) for the r- and $\theta$- components of neutrino flux. The left and right panels show profiles along the radial rays with $\theta = \pi/4$ and $3 \pi/4$, respectively. Only electron-type neutrinos with the local mean energy are considered in this figure. The time is $t=15$ms post bounce.
\label{signifikrth}} 
\end{figure*}
%%%%%%%%%%%%%%%%%%%%%%%%%%%%%%%%%

The existence of the non-axisymmetric features in the angular distributions of neutrinos and the appearance of the non-vanishing off-diagonal components of the Eddington tensor as a result are the main novel findings in this paper. The legitimate question then is how significant they are for supernova dynamics. In order to fully address this issue, it is required to run additional simulations with some approximate neutrino transport scheme such as the ray-by-ray and/or M1 methods, which either completely ignore or employ a makeshift prescription for these non-axisymmetric features, for the same progenitor, resolution, EOS and input physics and make a detailed comparison, which is certainly beyond the scope of this paper. Instead, in this section, we compare different components of the Eddington tensor quantitatively and discuss how the off-diagonal components might become important.

% In the following, however, we quantify the effect of off-diagonal components of the Eddington tensor based on our results including resolution studies, and then discuss the possible influence of supernova dynamics.

%on the time evolution of lower angular-moment of neutrinos

Note first that the equations for both the zeroth and first moments of the angular distribution include in principle all components of Eddington tensor (see, e.g., Eqs. (3.37) and (3.38) in \citet{2011PThPh.125.1255S}). It should be also pointed out that reaction rates of some neutrino-matter interactions such as non-isoenergetic scatterings and pair processes depend on higher-order moments including the Eddington tensor. The neglect of them may have some implications for CCSNe dynamics. Although this is an interesting issue and is in fact on our to-do-list, in the following, we will limit our discussion to the advection part of the neutrino transport.

% is limitted only for the transport of neutrinos. The influence via neutrino-matter interaction would be very important, which is actually one of our to-do-list in the future study.

%studied in the forthcoming paper.

%According to , the time evolution of first-neutrino angular moment (neutrino flux)

The principal part of the equations for the first angular moment or the flux can be approximately written as (see also Eq.~(3.38) in \citet{2011PThPh.125.1255S})
\begin{eqnarray}
&& \partial_t (F^r) \sim - \partial_r (E k^{r r}) - \frac{1}{r} \partial_{\theta} (Ek^{r \theta}) \label{eq:Frasymevo} , \\
&& \partial_t (F^{\theta}) \sim - \partial_r (E k^{r \theta}) - \frac{1}{r} \partial_{\theta} (E k^{\theta \theta}), \label{eq:Fthasymevo}
\end{eqnarray}
where we ignore collision terms and assume that the spacetime is flat and the background matter is axisymmetric and non-rotating. The off-diagonal component of Eddington tensor $k^{r \theta}$ appears in the second and first terms on the right hand side of Eqs.~(\ref{eq:Frasymevo}) and (\ref{eq:Fthasymevo}), respectively. Note that it does not show up in the principal part of the zeroth-order equation for the energy density.

In Fig.~\ref{signifikrth}, we display radial profiles of the absolute values of the ratios of $\partial_{\theta} (Ek^{r \theta})/r$ to $\partial_r (E k^{r r})$ (upper panels) and $\partial_r (E k^{r \theta})$ to $\partial_{\theta} (E k^{\theta \theta})/r$ (lower panels) on two radial rays with $\theta = \pi/4$ and $3\pi/4$ at $15$ms after bounce. In this analysis, we consider electron-type neutrinos alone, and their energy is set to the mean energy at each point. The results of both the normal- and high-resolution simulations are presented for comparison.

% The upper panels show the absolute ratio (2nd term/1st term) in Eq.~(\ref{eq:Frasymevo}) and the bottom panels show the ratio (1st term/2nd term) in Eq.~(\ref{eq:Fthasymevo}), both of which nominator are the relative term of $k^{r \theta}$. 

As seen in the upper panels, the radial flux is in general dictated mainly by $k^{r r}$ with $k^{r \theta}$ being at most $10\%$. This is certainly not a large value but still may not be ignored, since, as \citet{2016arXiv161105859B} claims, an accumulation of seemingly minor effects may turn out to be crucially important. On the other hand, $k^{r \theta}$ plays more important roles in the equation for the lateral component of neutrino flux as demonstrated in the bottom panels. In fact, the ratio of radial gradient of $E k^{r \theta}$ to the lateral gradient of $ Ek^{r \theta}/r$ exceeds unity in some post-shock regions. This is also the case for the result of the high-resolution simulation although the radial profiles themselves are quite different from those in the normal-resolution run, which is a consequence of the fact that matter distributions in the background become different between the two cases.

% Our high resolution simulation also possesses the value with same order of magnitude albeit different radial profiles.

% Note that, contray to the Eq.(\ref{eq:Frasymevo}), all terms are trivially zero in spherical symmetry, i.e., genuinely multi-dimensional origin.

 As discussed in sections~\ref{sec:distmom} and \ref{sec:resolution}, the M1 prescription is not very successful in reproducing $k^{r \theta}$ in the semi-transparent region particularly in non-spherical settings.
% Although there is no preferred direction in momentum space in the M1 method unlike the ray-by-ray approximation,
Although there is no artificially preferred direction in the M1 transport unlike in the ray-by-ray approximation, the lateral neutrino flux may be still inaccurate.
It is misleading to argue that the Eddington tensor is reproduced again very well in the transparent regime with its off-diagonal component becoming negligible compared with the dominant $k^{r r}$. This is because the errors in the semi-transparent region will not be confined there and spread to transparent region in time. The errors in the flux will lead to those in the Eddington tensor through the closure relation, which will again contribute to errors in the flux. This may eventually affect CCSNe dynamics. The quantitative assessment of this effect requires detailed comparisons in collaboration with other groups and is much beyond the scope of this first report of our new simulations.

% would be completely different from the multi-angle transport in reality. The influence will spread globally and change the profile of energy density of neutrinos, which is the fundamental quantities to dictate the neutrino absorption, and then influence CCSNe dynamics.

% meaning that all two-moment methods possess a similar risk as ray-by-ray approximations.

% that two-moment methods with the crrent closure prescription would possess the same extend of error (at least $\sim 10\%$) and risk as ray-by-ray approximation.

 It is finally mentioned that the above analysis is based on the result of the early post-bounce phase, in which the semi-transparent region is highly dynamical owing to the prompt convection, and the $k^{r \theta}$ effect may be much smaller in the later phase. The errors in early times have some influences on the evolution in later times in principle, though. It should be also added that convections in the proto-neutron star and other hydrodynamical instabilities such as SASI and convections in the heating region occur more often than not even in the late phase. It is repeated that the quantitative assessments are certainly in order and will be studied in subsequent papers.

\section{Comparison with previous works} \label{sec:compprev}
In this section, we attempt to make a comparison of our results with other CCSNe simulations. The same progenitor model has been employed by many authors so far \citep{2012ApJ...756...84M,2014ApJ...786...83T,2016ApJ...825....6S}. It is mentioned first that our results are qualitatively in line with them in that softer EOS's are advantageous for shock revival. It should be pointed out, however, that there are some studies, in which softer EOS including the LS EOS have smaller shock radii initially than the stiffer ones (see, e.g., \citet{2014EPJA...50...46F}), in apparent contradiction with our results.

 According to \citet{2014EPJA...50...46F}, the difference in the shock trajectory originates mainly not from the stiffness of EOS but from the treatment of electron captures on heavy nuclei: representative heavy nuclei tend to be smaller in the softer LS EOS than in the stiffer STOS EOS, which is essentially the same as our FS EOS except for the single-nucleus approximation in the former, resulting in the greater deleptonization in the LS EOS during the collapse phase; this in turn leads to the smaller inner core and hence the weaker prompt shock wave for the LS EOS. It should be recalled, however, that the electron capture rates employed in our simulation with the LS EOS are the same as those for the simulation with the FS EOS. As a result, the effects just mentioned are not taken into account in our current simulations and the shock trajectories reflect the difference in the stiffness of EOS's alone.

 The treatment of nuclear weak interactions consistent with the EOS employed is important to compute CCSNe dynamics accurately. We stress that the current approximate treatment is meant just for simplicity in models with EOS's that employ the single-nucleus approximation. We believe that multi-nucleus EOS's are indispensable for the quantitative study of the nuclear weak interactions mentioned above. Such a study indeed under way \citep{Nagakuraprep} with the multi-nucleus extension by \citet{2017PhRvC..95b5809F} of Togashi's EOS \citep{2013NuPhA.902...53T}, which is based on the variational method for realistic nulcear potentials.

It is also important to point out that the shock expansion in our model looks less energetic than those in other simulations with the same progenitor model (see e.g., \citet{2014ApJ...786...83T}). It is difficult to pin down the cause of the discrepancy, since there are many differences in input physics as well as numerical methods for hydrodynamics and neutrino transport, but the ray-by-ray approximation employed for neutrino transport in their simulations may be one of the main causes of the difference. In fact, \citet{2016ApJ...831...81S} pointed out that the ray-by-ray approximation tends to artificially facilitate explosion in 2D, enhancing sloshing motions in axisymmetry. A similar concern was also expressed by \citet{2015ApJS..216....5S}, in which they showed that the asymmetry in the neutrino heating tends to be overestimated in the ray-by-ray approximation. More detailed comparisons in collaborations with other groups are required to substantiate the claim, though.

\section{Summary and Discussion} \label{sec:summary}
We have presented the first report of spatially axisymmetric CCSNe simulations with the full Boltzmann neutrino transport. We have found both similarities and differences between the two models with two different nuclear EOS's. On the one hand, the neutrino luminosities and mean energies as well as the post-shock morphologies except the scale are very similar between the two. This seems to be a consequence of the cancellation of the stronger bounce that would be expected in the softer LS EOS by the greater electron captures that produced the smaller inner core in the LS EOS model. On the other hand, the neutrino-heating efficiency and the mass in the gain region are consistently higher for the LS EOS. This seems to be due to more vigorous turbulent motions in the post-shock flow for the LS EOS than for the FS EOS, the fact which results in the greater expansion of the shock wave: it has reached $\sim 700{\rm km}$ by $300{\rm ms}$ after bounce and its maximum radius is still growing.

 %Our results share the qualitative trend with previous studies \citep{2012ApJ...756...84M,2014ApJ...786...83T,2016ApJ...825....6S}, in which softer EOS's are advantageous for shock revival. It should be pointed out, however, that the shock expansion in our model looks less energetic than others (see, e.g.,~\citet{2016ApJ...818..123B}). It is also mentioned that the electron capture rates we employed for the LS EOS model are the same as those for the FS EOS and are not consistent with the EOS. Further studies are certainly needed to make clear what causes the difference in the shock dynamics.

% Although the progenitor we employed in this study produced explosions rather commonly in other approximate simulations~\citep{2012ApJ...756...84M,2014ApJ...786...83T,2016ApJ...825....6S} and our results share with them the qualitative trend that softer EOS's are advantageous for shock revival, it should be pointed out that {\bf the shock expansion in our model is not robust, which contradicts with some simulations (see e.g.,~\citet{2016ApJ...818..123B}).}

% the shock propagation after revival seems much less vigorous in our simulation than in others (see e.g.,~\citet{2016ApJ...818..123B}).

By virtue of the multi-angle treatment in our simulations, we have found interesting features in the neutrino distribution in momentum space, such as the lack of axisymmetry with respective to the local radial direction and the non-vanishing off-diagonal component of the Eddington tensor. With an aid of our previous analyses in \citet{2017ApJ...847..133R} and an additional high-resolution simulation for the early post-bounce phase, we have estimated that the current simulations may have underestimated the neutrino-heating rate by a few percent owing to rather low angular and energy resolutions in momentum space. The possible effects of the off-diagonal component of the Eddington tensor, $k^{r \theta}$, on neutrino transport have been also discussed quantitatively: it plays a non-negligible role for the time evolutions of neutrino fluxes; it may give a $\sim 10 \%$ level of contribution to the neutrino luminosity and, more importantly, can be a dominant factor for the time evolution of lateral flux in the semi-transparent region.

We have found an interesting correlation/anti-correlation in $k^{r \theta}$ between $\nu_e$ and $\bar{\nu}_e$ depending on the radius. It is related with the lateral fluxes of these neutrinos. It will be interesting to see how well the M1 approximation fares in reproducing these features and hence the lateral fluxes. The close comparison between our Boltzmann solver and other approximate methods possibly in collaboration with other groups will be indispensable to assess critically and quantitatively the significance of the findings in this paper for the CCSNe dynamics. It will also enable us to calibrate and possibly improve the prescriptions, which should be given by hand in approximate transport schemes. This is indeed important practically, since our method is very costly in terms of required numerical resources.

We have made an attempt to compare our results with those obtained by other groups for the same progenitor model. We have found that the general trend that softer EOS's are favorable for shock revival is also true of our simulations. On the other hand, the continuous shock expansion observed for the softer LS EOS looks less energetic than that found by others. Although this seems to be consistent with the finding by \citet{2016ApJ...831...81S} that the ray-by-ray approximation in spatial axisymmetry may artificially enhance shock revival, more detailed comparisons are certainly necessary to draw some conclusions.

% We need to compare our results with those obtained in other approximate simulations quantitatively more in detail. That will provide us with invaluable information that is not only indispensable to understand the origin of the differences in supernova dynamics mentioned above but will also enable us to calibrate and possibly improve the prescriptions, which should be set by hand in the approximate transport schemes. This is indeed important practically, since our method is very costly in terms of required numerical resources.

There are also certainly many other issues remaining to be addressed.
The top priority is to make detailed comparisons with other approximate methods to assess the importance of multi-angle treatments for supernova dynamics by possibly collaborating with other groups. We will also proceed to explore other progenitors with different masses. The EOS dependence should be further clarified. Rotation is another concern, since the angular distribution in momentum space is then qualitatively changed: e.g., the principal axis will not be aligned with the radial direction in general and another off-diagonal component, $k^{r \phi}$, will no longer be vanishing.
 We are currently implementing general relativity in our code to investigate its influences, which are expected to be non-negligible. The angular distributions for different species of neutrinos we obtained in this study are valuable in their own right for e.g. the analysis of collective oscillations of neutrino flavors \citep{2010ARNPS..60..569D,2013PhRvD..88g3004M,2017arXiv170603360C,2017PhRvL.118b1101I}, which feed on the differences in the angular distributions among different neutrino species. They are currently being investigated and the results will be reported elsewhere.
% We are the only ones at present that can provide realistic data required.
% Such an analysis is in fact being undertaken and the results will be published elsewhere. 

\acknowledgments

H.N. acknowledges to C. D. Ott, S. Richers, L. Roberts, D. Radice, M. Shibata, Y. Sekiguchi, K. Kiuchi and T. Takiwaki for valuable comments and discussions. The numerical computations were performed on K computer, at AICS, FX10 at Information Technology Center of Tokyo University, SR16000 at YITP of Kyoto University, and SR16000 and Blue Gene/Q at KEK under the support of its Large Scale Simulation Program (14/15-17, 15/16-08, 16/17-11), Research Center for Nuclear Physics (RCNP) at Osaka University, the XC30 and the general common use computer system at the Center for the Computational Astrophysics, CfCA, the National Astronomical Observatory of Japan. Large-scale storage of numerical data is supported by JLDG constructed over SINET4 of NII. H.N and S.F were supported in part by JSPS Postdoctoral Fellowships for Research Abroad No. 27-348 and 28-472 and H.N was partially supported at Caltech through NSF award No. TCAN AST-1333520. This work was supported by Grant-in-Aid for the Scientific Research from the Ministry of Education, Culture, Sports, Science and Technology (MEXT), Japan (15K05093, 24103006, 24105008, 24740165, 24244036, 25870099, 26104006, 16H03986, 17H06357, 17H06365) and HPCI Strategic Program of Japanese MEXT and K-computer at the RIKEN and Post-K project (Project ID: hp 140211, 150225, 160071, 160211, 170230, 170031, 170304).

\end{document}